\documentclass [12pt]{article}
\usepackage{amssymb,amsfonts,amsmath,bm}

\usepackage{graphicx}

\usepackage{psfrag}

\usepackage{setspace}

\usepackage{subfigure}

\date{}

\begin{document}

\begin{flushleft}
{\Large
\textbf{Electrokinetic instability in the sharp interface limit: The perpendicular electric field case}
}
\\
H.-Y. Hsu and Neelesh A. Patankar$^{\ast}$\\

Department of Mechanical Engineering, Northwestern University, 
\\
2145 Sheridan Road, Evanston, IL 60208, USA

$\ast$ Corresponding author (n-patankar@northwestern.edu)
\end{flushleft}




\section*{Abstract}

In this paper, instability at an interface between two miscible
liquids with identical mechanical properties but different
electrical conductivities is analyzed in the presence of an
electric field that is perpendicular to the interface. A parallel
electric field case was considered in a previous work
\cite{pata11a}. A sharp Eulerian interface is considered between
the two miscible liquids. Linear stability analysis leads to an
analytic solution for the critical condition of instability. The
mechanism of instability is analyzed. Key differences between the
perpendicular and parallel electric field cases are discussed.
The effect of a microchannel geometry is studied and the relevant
non-dimensional parameters are identified.

\section{Introduction}

Instabilities at an interface between two miscible liquids with
identical mechanical properties but different electrical
conductivities, in the presence of externally applied electric
fields, have been studied by Santiago and co-workers
\cite{oddy01a,linh04a,oddy05a,chen05a,stor05a,posn06a,linh08a}.
These are instabilities in strong electrolytes and have been
termed electrokinetic instabilities. Instabilities with leaky
dielectrics have also been considered in literature \cite{uguz08a,
uguz08b}. In this paper, electrokinetic instabilities are
considered where the applied electric field is perpendicular to
the interface between two strong electrolytes. This case is
expected to be more unstable compared to the parallel electric
field case \cite{chen05a}.

In prior analytic work
\cite{linh04a,oddy05a,chen05a,stor05a,posn06a,boyd07a,linh08a}, a
diffuse interface was considered between the two miscible
electrolytes. Following the approach in our earlier work
\cite{pata11a}, we show that the correct behavior in the
perpendicular electric field configuration can be also obtained
with a sharp Eulerian interface between the two miscible
electrolytes. A diffuse interface is not required. The assumption
of a sharp interface leads to an easier analytic problem, which
results in a compact non-dimensional parameter that determines the
unstable behavior of the system. In the above problem, the
unstable behavior is quantitatively influenced by the thickness of
the diffuse interface between the two liquids. However, the sharp
interface case, which corresponds to an experiment where the
electric field is applied before the interface has diffused
significantly, is an important limiting case.

Although the approach used in this paper is same as that used by
Patankar \cite{pata11a}, the difference in orientation of the
applied electric field brings out different parametric behavior,
e.g., the critical condition for instability. The difference in
the parametric dependence cannot be intuitively deduced
(especially the dependence on electrical conductivities) and is
difficult to obtain simply based on experimental data.

In the following sections, the governing equations will be
presented first. Instabilities in two geometric configurations --
an infinite domain and a microchannel geometry -- will be studied.

\section{Infinite domain}
\subsection{Problem formulation}

The interface is an Eulerian surface which is defined with respect
to the base state. The electrical conductivity in the base state
changes sharply at this Eulerian interface. An external electric
field $\mathbf{E}$ is applied in the $y$-direction, which is
perpendicular to the interface. Liquid $a$ is above the interface
(positive values of $y$), and liquid $b$ is below it (negative
values of $y$). In the first configuration considered here, the
domain is infinite.

The governing equations for this problem can be summarized as
follows \cite{pata11a}:
\begin{equation}
\left.
\begin{array}{l}
\varepsilon\nabla^2\phi=-\rho_b,\\
\nabla\cdot(\sigma\mathbf{E})=0,\\
\frac{\displaystyle D \sigma}{\displaystyle D t}=D_\sigma\nabla^2\sigma,\\
\nabla\cdot\mathbf{u} = 0,
\rho\frac{\displaystyle\partial\mathbf{u}}{\displaystyle\partial
t} + \rho (\mathbf{u}\cdot\nabla)\mathbf{u} = -\nabla
p+\mu\nabla^2\mathbf{u}+\rho_b\mathbf{E},
\end{array}
\right\}\label{eqn:governingeqn}
\end{equation}
where $\varepsilon$ is the permittivity, $\phi$ is the electric
potential, $\rho_b$ is the bulk charge density in the liquids,
$\sigma$ is the electrical conductivity which is different in
liquids a and b, $\mathbf{E}$ is the electrical field, $D_\sigma$
is the diffusion coefficient for the electrical conductivity,
$\mathbf{u}$ is the velocity field, $\rho$ is the density, $p$ is
the dynamic pressure (gravity is balanced by the hydrostatic
component), and $\mu$ is the viscosity. Liquids $a$ and $b$ are
assumed to be strong electrolytes which implies that the
electrical conductivities are high. A binary electrolyte is
considered \cite{pata11a}.

The jump conditions, at the Eulerian interface defined above, are
summarized below \cite{pata11a}
\begin{eqnarray}
\rho_s=\|\varepsilon\mathbf{E}\|\cdot\mathbf{n}\label{eqn:surfacecharge2},\\
\|\sigma\mathbf{E}\|\cdot\mathbf{n}=0,\label{eqn:chargecontinuity}\\
\|\sigma\mathbf{u}-D_\sigma\nabla\sigma\|\cdot\mathbf{n}=0\label{eqn:diffusioncharge2},\\
\|\mathbf{u}\|\cdot\mathbf{n}=0\label{eqn:vcontinuity},\\
\mathbf{n}\cdot\|\mathbf{\tau}-\rho\mathbf{u}\mathbf{u}\|\cdot\mathbf{n}=0,\mbox{and}\hspace{0.2cm}
\mathbf{t}\cdot\|\mathbf{\tau}-\rho\mathbf{u}\mathbf{u}\|\cdot\mathbf{n}=0,\label{eqn:stressinterface}
\end{eqnarray}
where
$\displaystyle\mathbf{\tau}=-p\mathbf{I}+\mu(\nabla\mathbf{u}+
\nabla\mathbf{u}^T)+
\varepsilon\mathbf{E}\mathbf{E}-\frac{\varepsilon}{2}
\mathbf{E}\cdot\mathbf{E}\mathbf{I}$, and $\|\|$ denotes the value
of the variables at the interface in liquid a minus the value in
liquid b. $\mathbf{n}$ is a unit normal to the interface pointing
into liquid a.

Semi-infinite domains are considered here in both liquids $a$ and
$b$. The sharp Eulerian interface is located at the center of the
domain ($y = 0$). The material properties $\rho$, $\mu$, and
$\varepsilon$ are assumed to be same and constant in both liquids
$a$ and $b$. Only the electrical conductivities are considered to
be different in liquids $a$ and $b$. Since the domain is unbounded
and symmetric with respect to the $z$-direction, the problem is
considered to be two-dimensional in the $x$-$y$ plane. It is
assumed that there is no electroosmotic flow in the base state.
This assumption is discussed further in the Discussion section.
The base solution is given by
\begin{equation}
\left.
\begin{array}{l}
\mathbf{u_0}=\mathbf{0},\\
\rho_{b0}=0,\\
\sigma_0=\sigma_0^{a} \hspace{0.5cm}\mbox{in liquid $a$} ,\\
\sigma_0=\sigma_0^{b}\hspace{0.5cm}\mbox{in liquid $b$} ,\\
\displaystyle\mathbf{E_0}^a=\frac{I_0}{\sigma_0^a}\mathbf{j},\hspace{1cm}\displaystyle\mathbf{E_0}^b=\frac{I_0}{\sigma_0^b}\mathbf{j},\\
\displaystyle\rho_{s0}=\frac{\varepsilon_0I_0\Delta\sigma_0}{\sigma_0^{a}\sigma_0^{b}},\\
\displaystyle p_0^a=\frac{\varepsilon I_0^2}{2(\sigma_0^{a})^2},\hspace{1cm}\displaystyle p_0^b=\frac{\varepsilon I_0^2}{2(\sigma_0^{b})^2},\\
\label{eqn:basesol}
\end{array}
\right\}
\end{equation}
where $\rho_{b0}$ is the bulk free charge per unit volume inside
the fluid, $\rho_{s0}$ is the charge per unit area at the
interface, $\Delta\sigma_0=\sigma_0^a-\sigma_0^b$, and subscript
$0$ denotes the variables in the base state. $I_0$ is a constant
current in $y$-direction. Perturbations are superimposed on the
base solution. The conductivity profile in the base state will
diffuse with time. However, an approximation is introduced by
assuming that the conductivity profile is ``frozen" with a sharp
jump at the interface. This is not a fully consistent
approximation but it is found to be reasonable when the time scale
of the instability is short \cite{chen05a,linh04a}.

After linearization, the governing equations, under the assumption
of a ``frozen" base state, for the perturbations are

\begin{equation}
\left.
\begin{array}{l}
\nabla\cdot\mathbf{u'} = 0,\\
\displaystyle\varepsilon\nabla^2\phi'=-\rho_b',\\\displaystyle
\sigma_0\nabla^2\phi'-\frac{I_0}{\sigma_0}\frac{\partial\sigma'}{\partial
y}=0,\\
\displaystyle \frac{\partial \sigma'}{\partial t}=D_\sigma\nabla^2\sigma',\\
\displaystyle\rho\frac{\partial\mathbf{u'}}{\partial t}= -\nabla
p'+\mu\nabla^2\mathbf{u'}+\rho_b'\mathbf{E_0},
\label{eqn:perteqn}
\end{array}
\right\}
\end{equation}
where superscript $'$ denotes perturbations. The dimensional form
of the perturbations is given by
\begin{equation}
\left(
\begin{array}{l}
\mathbf{u}'\\
p'\\
\phi'\\
\rho_b'\\
\sigma'\\\
v_{\mbox{int}}'\\
\end{array}
\right )\mbox{=} \left(
\begin{array}{l}
 \mathbf{u}(y)\\
p(y)\\
 \phi(y)\\
\rho_b(y)\\
\sigma(y)\\
 v_{\mbox{int}}
\end{array}
\right ) e^{-ikx+st},\label{eqn:solnperturb}
\end{equation}
where $v_{\mbox{int}}$ is the $y$ velocity component at the
interface, $k$ (a real number) is the wave number of the
perturbation, $s$ (a complex number) is the growth rate, and
$\mathbf{u}$, $p$, $\phi$, $\rho_b$, and $\sigma$ are the
amplitudes of the perturbations. An instability is implied by a
positive real part of $s$.

The governing equations are non-dimensionalized by using the
following scales
\begin{eqnarray}
Length\rightarrow H,  \phi'\rightarrow\frac{I_0H}{\sigma_m},
 E'\rightarrow\frac{I_0}{\sigma_m},
 \rho_b'\rightarrow\frac{\varepsilon I_0}{H\sigma_m}, p'\rightarrow\frac{\varepsilon I_0^2}{\sigma_m^2}
\nonumber\\
\sigma'\rightarrow\sigma_m(=\frac{(\sigma_0^a+\sigma_0^b)}{2}),
 Velocity\rightarrow\frac{H \varepsilon I_0^2}{\sigma_m^2\mu},
 time\rightarrow\frac{\mu \sigma_m^2}{\varepsilon
I_0^2}.\label{eqn:NDparameter}
\end{eqnarray}

The length scale is $H$ = 1/$k$. The velocity scale is based on
the balance between the viscous and electrical forces in the
momentum equation. After non-dimensionalization, the perturbation
equations become
\begin{equation}
\left.
\begin{array}{l}
\nabla\cdot\mathbf{u'} = 0,\\
\nabla^2\phi' =-\rho_b', \\
\displaystyle\sigma_{0N}^2\nabla^2\phi'-\frac{\partial\sigma'}{\partial y} =0,\\
\displaystyle Pe\frac{\partial{\sigma'}}{\partial{t}}=\nabla^2\sigma',\\
\displaystyle Re\frac{\partial\mathbf{u'}}{\partial t}= -\nabla
p'+\nabla^2\mathbf{u'}+\frac{\rho_b'}{\sigma_{ON}}\mathbf{j}.\label{eqn:infinitendperturbation}
\end{array}
\right\}
\end{equation}
where same symbols have been retained for the non-dimensional
variables. The non-dimensional parameters in the governing
equations are
\begin{equation}
\left.
\begin{array}{r}
\displaystyle Re=\frac{\rho\varepsilon I_0^2}{\mu^2\sigma_m^2k^2},\\
\displaystyle Pe=\frac{\varepsilon I_0^2}{\mu D_\sigma \sigma_m^2 k^2},\\
\displaystyle\sigma_{0N}=\frac{2\sigma_0}{\sigma_0^{a}+\sigma_0^{b}},
\end{array}
\right\}
\end{equation} where $Re$ is the Reynold's number, $Pe$ is the
Peclet number which is a ratio of convection and diffusion terms,
and $\sigma_{0N}$ is the non-dimensionalized value of $\sigma_0$
in each liquid.

In the non-dimensional form of Equation (\ref{eqn:solnperturb}),
we put $k$ = 1 since the length is non-dimensionalized by 1/$k$,
and $s$ will be understood to be non-dimensionalized by the
inverse of the time scale. Inserting the non-dimensional form of
Equation (\ref{eqn:solnperturb}) into Equation
(\ref{eqn:infinitendperturbation}) and simplifying we get
\begin{equation}
\left.
\begin{array}{l}
\displaystyle(D^2-1-Pes)\sigma=0,\\
\displaystyle(D^2-1)\phi=\frac{D\sigma}{\sigma_{ON}^2},\\
\displaystyle(D^2-1-Res)(D^2-1)\upsilon=-\frac{D\sigma}{\sigma_{ON}^3},\\
u=-iD\upsilon,\\
P=(D^2-1-Res){Dv}\label{eqn:different},
\end{array}
\right\}
\end{equation}
where $s$ is the non-dimensional growth rate, $D$ is the
derivative with respect to $y$, and $u$, $v$ are the $x$, $y$
components of velocity, respectively. The solution of Equation
(\ref{eqn:different}) should approach zero as
$y\rightarrow\pm\infty$ in liquids $a$ and $b$, respectively. This
gives the following solutions for liquids $a$ and $b$
\begin{equation}
\left.
\begin{array}{l}
\sigma^a=A^ae^{-\lambda y},\\
\displaystyle\phi^a=B^ae^{-y}+\frac{D\sigma^a}{(\sigma_{0N}^a)^2(Pes)},\\
\displaystyle v^a=C^ae^{-y} +D^a e^{-q y}-\frac{D\sigma^a}{(Pes-Res)(Pes)(\sigma_{0N}^a)^3},\\
\end{array}
\right\}\label{eqn:liquida}
\end{equation}
\begin{equation}
\left.
\begin{array}{l}
\sigma^b=A^be^{\lambda y},\\
\displaystyle\phi^b=B^b e^{y}+\frac{D\sigma^b}{(\sigma_{0N}^b)^2(Pes)},\\
\displaystyle v^b=C^be^{y} +D^be^{q y}-\frac{D\sigma^b}{(Pes-Res)(Pes)(\sigma_{0N}^b)^3},\\
\end{array}
\right\}\label{eqn:liquidb}
\end{equation}
where $\lambda$ and $q$ are positive and are given by
\begin{equation}
\left.
\begin{array}{r}
\lambda=\sqrt{1+Pes},\\
q=\sqrt{1+Res}.
\end{array}
\right\}
\end{equation}
Superscripts $a$ or $b$ denote liquids $a$ or $b$, respectively.
$A, B, C$ and $D$'s with superscript are constants.

The linearized jump conditions for the electrical conductivity at
the interface are given by
\begin{equation}
\left.
\begin{array}{r}
\|\sigma\|=0,\\
\|D\sigma\|=Pev_{\mbox{int}}\Delta\sigma_{ON},
\end{array}
\right\} \label{eqn:jumpconductivity}
\end{equation}
where $\displaystyle v_{\mbox{int}} = v^a = v^b =
\frac{(v^a+v^b)}{2}$  at $y$ = 0 is the amplitude of the
non-dimensional perturbation velocity at the interface, and
$\displaystyle\Delta\sigma_{ON}=\sigma_{ON}^a-\sigma_{ON}^b$. The
jump conditions in Equation (\ref{eqn:jumpconductivity}) follow by
assuming that there is no self-sharpening mechanism that creates
discontinuities in the electrical conductivity. This is
reasonable, since the diffusive behavior of $\sigma$ is important
in this problem \cite{Melcher}. The second jump condition in
Equation (\ref{eqn:jumpconductivity}) follows from Equation
(\ref{eqn:diffusioncharge2}). Using these conditions, we get the
solution for $A^a$ and $A^b$ as
\begin{equation}
A^a=A^b=-\frac{Pev_{\mbox{int}}\Delta\sigma_{ON}}{2\lambda}.\label{eqn:answerA}
\end{equation}
To solve for $B^a$ and $B^b$, we use the interface jump conditions
for the electric potential:
\begin{equation}
\left.
\begin{array}{r}
\|\phi\|=0,\\
\displaystyle\|\sigma_{ON}D\phi\|+\sigma_{y=0}\frac{\Delta\sigma_{0N}}{{\sigma_{0N}^a\sigma_{0N}^b}}=0.
\end{array}
\right\} \label{eqn:jumpelectricpot}
\end{equation}
The first jump condition in Equation (\ref{eqn:jumpelectricpot})
follows by assuming that no double layers are formed at the
interface. This is reasonable since the interface is not
insulating and it is assumed that the current carrying species can
pass from one side of the interface to the other \cite{Melcher}.
The second jump condition in Equation (\ref{eqn:jumpelectricpot})
follows from Equation (\ref{eqn:chargecontinuity}). Thus, $B^a$
and $B^b$ are
\begin{eqnarray}
B^a=\frac{Pev_{\mbox{int}}\triangle\sigma_{0N}^2}{4\sigma_{0N}^a\sigma_{0N}^b(\lambda^2-1)\lambda}
-\frac{Pev_{\mbox{int}}\triangle\sigma_{0N}(\sigma_{0N}^{a2}+\sigma_{0N}^{b2})}{4(\lambda^2-1)\sigma_{0N}^a\sigma_{0N}^{b2}},\\
B^b=\frac{Pev_{\mbox{int}}\triangle\sigma_{0N}^2}{4\sigma_{0N}^a\sigma_{0N}^b(\lambda^2-1)\lambda}
+\frac{Pev_{\mbox{int}}\triangle\sigma_{0N}(\sigma_{0N}^{a2}+\sigma_{0N}^{b2})}{4(\lambda^2-1)\sigma_{0N}^{a2}\sigma_{0N}^b}.
\end{eqnarray}
The linearized jump conditions for the velocity (from Equation
\ref{eqn:vcontinuity}) and the stress (from Equation
\ref{eqn:stressinterface}) at $y$ = 0 give
\begin{equation}
\left.
\begin{array}{r}
\|v\|=0,\\
\|u\|=0,\\
\displaystyle\|-p+2Dv-\frac{D\phi}{\sigma_{0N}}\|=0,\\
\displaystyle\|D^2v+v-\frac{\phi}{\sigma_{0N}}\|=0.\label{eqn:jcforv}
\end{array}
\right\}
\end{equation}
Note that the jump condition for $u$ is due to the no-slip
condition. This is reasonable since no double-layers are formed at
the interface. Inserting the solutions for velocity in Equations
(\ref{eqn:liquida}) and (\ref{eqn:liquidb}) into Equation
(\ref{eqn:jcforv}), we get five homogeneous equations for five
unknowns: $C^a$, $C^b$, $D^a$, $D^b$, and $v_{\mbox{int}}$:
\begin{equation}
\left(
\begin{array}{ccccc}
1 & -1 &1 & -1 & f_1\\
1 & 1 & q & q & f_2\\
1 & 1 & q^3 & q^3 & f_3\\
1 & -1 & q & q^2 & f_4\\
1 & 1 & 1 & 1 & f_5\\
\end{array}
\right ) \left(
\begin{array}{l}
 C^a\\
 C^b\\
 D^a\\
 D^b\\
 v_{\mbox{int}}
\end{array}
\right )\mbox{=} \left (
\begin{array}{l}
 0\\
 0\\
 0\\
 0\\
 0
 \end{array}
 \right )
\end{equation}

\begin{eqnarray}
f_1&=&-\frac{Pe\Sigma(\Gamma-2)}{2(\lambda^2-1)(\lambda^2-q^2)}\sqrt{\frac{\Gamma-3}{\Gamma+1}},\\
f_2&=&\frac{\lambda Pe\Sigma\Gamma}{2(\lambda^2-1)(\lambda^2-q^2)},\\
f_3&=&\frac{\lambda^3Pe\Gamma}{2(\lambda^2-1)(\lambda^2-q^2)}
+\frac{Pe\Gamma}{2(\lambda^2-1)}(\frac{1}{\lambda}+(\Gamma-1)-\lambda\Gamma),\\
f_4&=&\frac{\lambda^2Pe\Sigma(\Gamma-2)}{2(\lambda^2-1)(\lambda^2-q^2)}\sqrt{\frac{\Gamma+1}{\Gamma-3}}
+\frac{Pe\Gamma}{2(\lambda^2-1)}\sqrt{\frac{\Gamma-3}{\Gamma+1}}\nonumber\\
&+&\frac{Pe\Sigma(\Gamma-1)^2}{2(\lambda^2-1)\sqrt{(\Gamma+1)(\Gamma-3)}}
-\frac{Pe\Sigma(\Gamma-2)}{2(\lambda^2-1)}\sqrt{\frac{\Gamma+1}{\Gamma-3}},\\
f_5&=&\frac{Pe\Sigma\Gamma}{2(\lambda^2-1)(\lambda^2-q^2)}-2.
\end{eqnarray}
where
$\displaystyle\Sigma=\frac{\Delta\sigma_{0N}^2}{\sigma_{0N}^{a2}\sigma_{0N}^{b2}}$
and
$\displaystyle\Gamma=\frac{\sigma_{0N}^a}{\sigma_{0N}^b}+\frac{\sigma_{0N}^b}{\sigma_{0N}^a}+1$.
The dispersion equation is obtained by setting the determinant of
the matrix to zero.

\subsection{The critical condition for instability}

The dispersion equation obtained from manipulations in Mathematica
is
\begin{eqnarray}
-(4\Gamma(1+\sqrt{1+ST})((1+S)^{\frac{3}{2}}\sqrt{1+ST}+\nonumber\\
\sqrt{1+S}(1+ST)+(1+S)\sqrt{1+ST}(1+\sqrt{1+ST})))\nonumber\\
+P_{\Sigma}(\sqrt{1+S}-\Gamma\sqrt{1+S}+\sqrt{1+ST})=0,
\end{eqnarray}
where $S=Pes$, $P_\Sigma=Pe\Sigma\Gamma$, and $\displaystyle
T=\frac{Re}{Pe}$. The maximum growth rate for this problem is the
largest root of the dispersion equation. It is verified that the
maximum growth rate is positive and real.

\begin{figure}
\centering
\includegraphics[width=13cm]{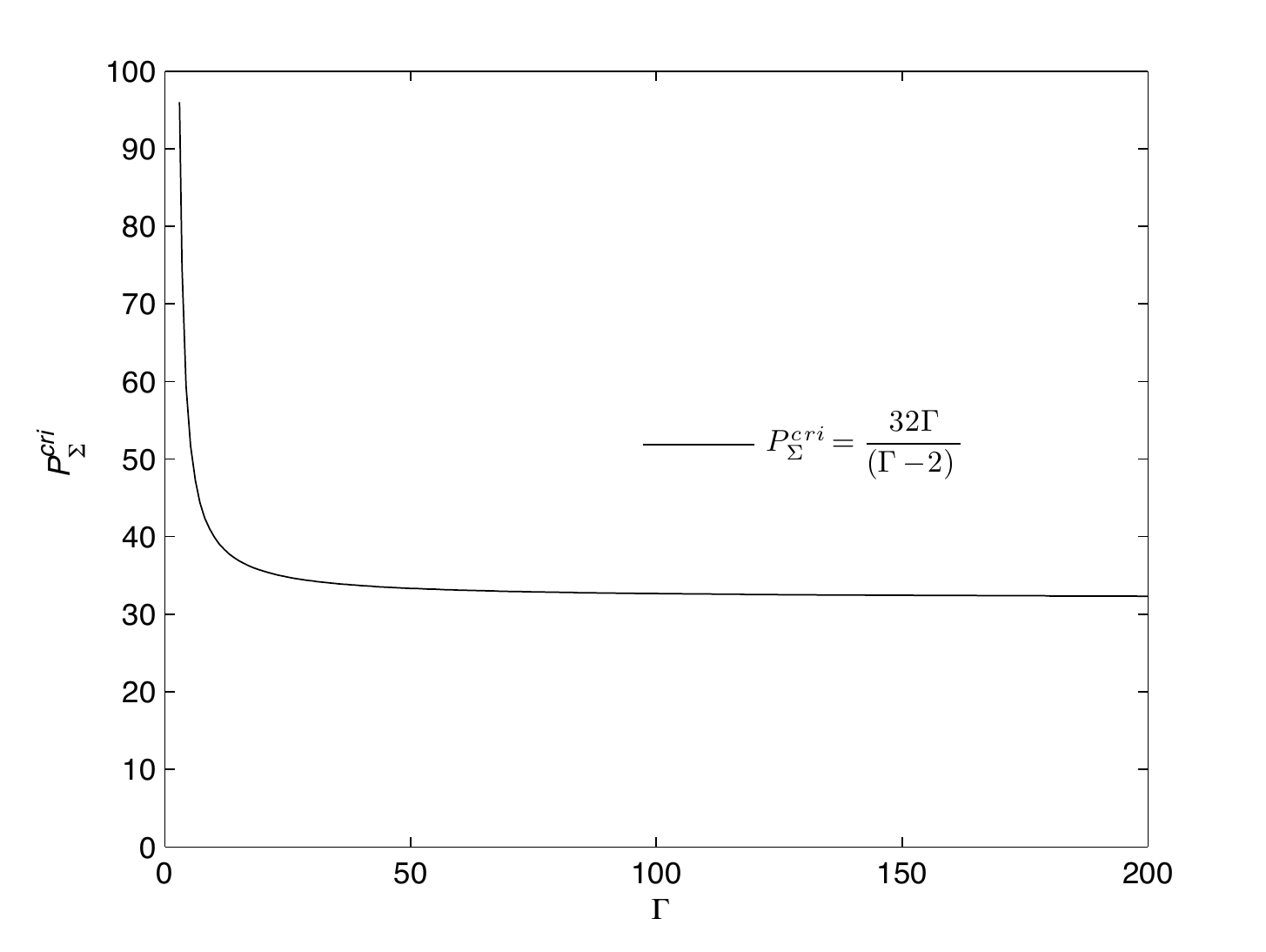}
\caption{The marginal stability curve ($P_\Sigma^{cri}$ vs.
$\Gamma$) identifying the critical condition for the onset of
instability. The region above the curve indicates unstable
conditions.} \label{fig:psigmavsgaminfinite}
\end{figure}

When the system is marginally stable i.e. when the maximum growth
rate $S=0$, then $P_\Sigma$ is found to be independent of $T$ and
is a function of $\Gamma$. The marginal stability condition gives
the critical condition for the onset of instability. It is given
by
\begin{equation}
\left.
\begin{array}{r}
\displaystyle P_\Sigma^{cri}=\frac{32\Gamma}{\Gamma-2},\\
\displaystyle\Rightarrow(\frac{\varepsilon
{E_{\mbox{mean}}}^2}{D_\sigma\mu
k^2})^{cri}=\frac{32}{(\Gamma-2)\Sigma},\label{eqn:pcri}
\end{array}
\right\}
\end{equation}
where $\displaystyle E_{\mbox{mean}}=\frac{I_0}{\sigma_{m}}$.
Figure (\ref{fig:psigmavsgaminfinite}) shows a plot of
$P_\Sigma^{cri}$ as a function of $\Gamma$.

\subsection{The mechanism of instability }

\begin{figure}
\centering
\includegraphics[width=8cm]{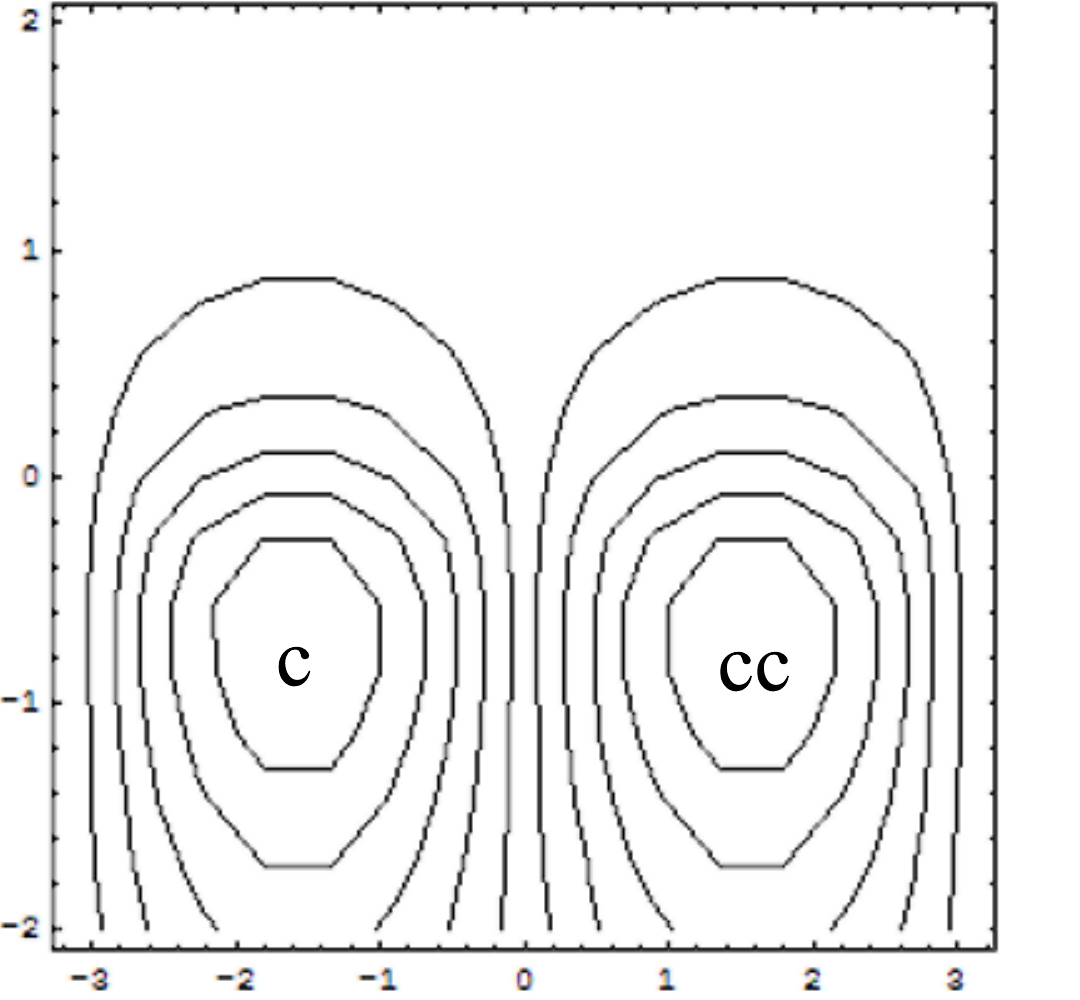}
\caption{Streamlines of an unstable mode in an infinite domain
where the applied electric field is perpendicular to the interface
between the two liquids. 'c' denotes clockwise rotation of the
fluid and 'cc' denotes counterclockwise rotation.}
\label{fig:streamline}
\end{figure}

An approach similar to that in our earlier work \cite{pata11a} is
followed. Streamlines spanning one wavelength are plotted for a
typical unstable mode in Figure (\ref{fig:streamline}).
The parameters are $T=0.05$, $S=25$, $P_\Sigma=683.3914$,
$\displaystyle\frac{\sigma_{0N}^a}{\sigma_{0N}^b}=10$ , and
$\Gamma=11.1$.

\begin{figure}
\centering
\includegraphics[width=13cm]{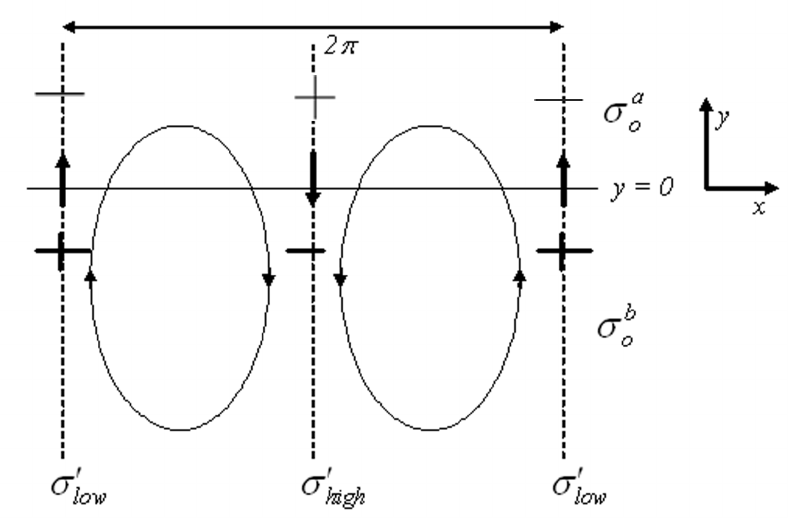}
\caption{A typical unstable flow cell is depicted. Higher
conductivity fluid is in the upper half and the lower conductivity
fluid is in the lower half. Interfacial velocity perturbations are
shown by bold arrows. Positive and negative signs show the
locations of high and low values of the perturbed electrical
conductivity. Bold signs indicate that there is greater bulk
charge in the lower conductivity fluid. } \label{fig:mech}
\end{figure}

A perturbation in the interfacial velocity,
$v'_{\mbox{int}}=cos(x)$, leads to a perturbation in the
electrical conductivity due to the electrohydrodynamic coupling in
Equation (\ref{eqn:jumpconductivity}), and Equations
(\ref{eqn:liquida}), (\ref{eqn:liquidb}) and (\ref{eqn:answerA}):
\begin{equation}
\sigma'^a=-\frac{Pe\Delta\displaystyle{\sigma_{ON}}}{2\lambda}e^{-\lambda
y}cos(x),\hspace{1cm}
\sigma'^b=-\frac{Pe\Delta\displaystyle{\sigma_{ON}}}{2\lambda}e^{\lambda
y}cos(x). \label{eqn:checksigma}
\end{equation}
It follows from Equation (\ref{eqn:checksigma}) that a region of
lower electrical conductivity is formed when $v_{\mbox{int}}$ is
maximum, and higher electrical conductivity is formed when
$v_{\mbox{int}}$ is minimum. This is depicted in Figure
(\ref{fig:mech}) with vertical bold arrows at locations of maximum
and minimum velocities. Perturbations in the electrical
conductivity leads to a perturbation in the bulk charge density
(Equation \ref{eqn:infinitendperturbation}) according to
$\rho_b'=-\frac{1}{\displaystyle{\sigma_{ON}^2}}
\displaystyle{\frac{\partial\sigma'}{\partial y}}$. Thus, we get
\begin{equation}
\rho_b'^a=-\frac{Pe\Delta\displaystyle{\sigma_{ON}}}{2\displaystyle{\sigma_{ON}^{a2}}}
e^{-\lambda y}cos(x),
\hspace{1cm}\rho_b'^b=\frac{Pe\Delta\displaystyle{\sigma_{ON}}}
{2\displaystyle{\sigma_{ON}^{b2}}}e^{\lambda y}cos(x).
\label{eqn:checkbulkcharge}
\end{equation}

This leads to an asymmetric bulk charge distribution in the domain
as seen in Figure (\ref{fig:mech}). The consequent electrical body
force in the fluid gives rise to a cellular flow that reinforces
the initial perturbation in velocity and causes instability.

\subsection{Comparison with the parallel electric field case}

It has been reported that when an electric field is applied
perpendicular to the interface the system is more unstable
compared to the parallel electric field case \cite{oddy01a}.

We consider this issue by comparing the critical condition for
instability for these cases. The following critical condition for
instability of the parallel electric field case is given in our
previous work \cite{pata11a}:
\begin{equation}
\left.
\begin{array}{r}
\displaystyle P_{\Sigma\parallel}^{cri}=32,\\
\displaystyle\Rightarrow(\frac{\varepsilon {E_0}^2}{D_\sigma\mu
k^2})^{cri}=\frac{32}{\Sigma_\parallel},\label{eqn:pcriparallel}
\end{array}
\right\}
\end{equation} where $\displaystyle\Sigma_\parallel=
\frac{\Delta\sigma_{0N}^2}{\sigma_{0N}^{a}\sigma_{0N}^{b}}.$
Comparison between Equations (\ref{eqn:pcri}) and
(\ref{eqn:pcriparallel}) implies that the perpendicular electric
field case is more unstable compared to the parallel electric
field case. This is discussed below.

Consider parameters: $\sigma^a/\sigma^b=10$ , $\mu=0.001 kg/ms$,
$\varepsilon=6.9\times10^{-10}C/Vm$, $\rho=1000kg/m^3$ , and
$D_{\sigma}=10^{-9}m^2/s$. We express $\Sigma$'s in terms of
$\sigma_a$ and $\sigma_b$ rather than $\sigma_{0N}^{a}$ and
$\sigma_{0N}^{b}$ in both parallel and perpendicular electric
field cases, Thus
\begin{equation}
\Sigma_\parallel=\frac{(\sigma^a-\sigma^b)^2}{\sigma^a\sigma^b}=8.1.
\end{equation}
Using equation (\ref{eqn:pcriparallel}), we get
\begin{equation}
\displaystyle(\frac{\varepsilon {E_0}^2}{D_\sigma\mu
k^2})^{cri}=\frac{32}{8.1}\approx 3.95.\label{eqn:cri1}
\end{equation}
For the perpendicular electric field case, we have
\begin{equation}
\left.
\begin {array}{r}
\Sigma=\displaystyle{(\frac{(\sigma^a)^2-(\sigma^b)^2}{2\sigma^a\sigma^b})^2}=24.5025,\\
\Gamma=\displaystyle{\frac{\sigma^a}{\sigma^b}+\frac{\sigma^b}{\sigma^a}}+1=11.1.
\end{array}
\right\}
\end{equation}
Using equation (\ref{eqn:pcri}), we get
\begin{equation}
\displaystyle(\frac{\varepsilon{E_{\mbox{mean}}}^2}{D_\sigma\mu
k^2})^{cri}=
\frac{32}{\Sigma(\Gamma-2)}=\frac{32}{24.5025\times9.1}\approx0.143.\label{eqn:cri2}
\end{equation}
Comparing Equations (\ref{eqn:cri1}) and (\ref{eqn:cri2}), we see
that the electric field needed to achieve the instability in the
parallel case is larger than the perpendicular electric field
case.

This difference is due to the difference in the flow pattern in
the unstable modes. The parallel electric field case has a
different cellular flow pattern in the unstable mode (see
\cite{pata11a}). Two pairs of counter rotating vortices are
produced in the parallel electric field case. This flow pattern is
much less asymmetric and gives rise to weaker flows. In the
perpendicular electric field case the asymmetry is much stronger
primarily due to the current flowing perpendicular to the
interface in the base state. This results in stronger
destabilizing forces in the perpendicular electric field case thus
making it more unstable.

We consider a single parameter above simply for the purpose of
comparing parallel and perpendicular electric field cases at
typical conditions that are known. Otherwise, the critical
conditions for the two cases (parallel and perpendicular) allow
comparison over the entire parameter range which shows a similar
trend that the perpendicular electric field case is more unstable.

\section{Shallow channel}
\subsection{Problem formulation}

\begin{figure}
\centering
\includegraphics[width=9cm]{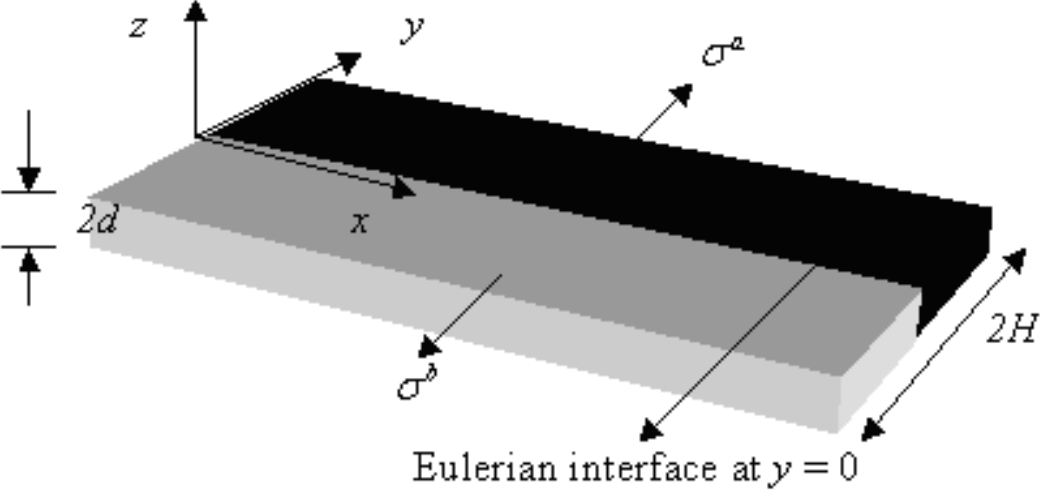}
\caption{The shallow channel geometry. }
\label{fig:shallowchannel}
\end{figure}

Next we apply the linear stability analysis to the case of a
shallow microchannel geometry (Figure \ref{fig:shallowchannel})
that is typical in microfluidic devices \cite{oddy01a,chen05a}.
The objective is to understand the influence of the device
geometry on the instability.

An external electrical field applied perpendicular to the
interface ($y$-direction) between two liquids $a$ and $b$ in a
domain that is unbounded in the $x$ direction. It is assumed that
there is no electroosmotic flow, and there is no charges in the
base state. The base state is the same as that in Equation
(\ref{eqn:basesol}) and the perturbation equations are given by
Equation (\ref{eqn:perteqn}).

The governing equations are non-dimensionalized by using the
following scales
\begin{eqnarray}
x,y\rightarrow H, z\rightarrow d,
\phi'\rightarrow\frac{I_0H}{\sigma_m},
 E'\rightarrow\frac{I_0}{\sigma_m},
 \rho_b'\rightarrow\frac{\varepsilon I_0}{H\sigma_m}, p'\rightarrow\frac{\varepsilon I_0^2}{\sigma_m^2},
\nonumber\\
\sigma'\rightarrow\sigma_m(=\frac{(\sigma_0^a+\sigma_0^b)}{2}),
 Velocity\rightarrow\frac{H \beta^2 \varepsilon I_0^2}{\sigma_m^2\mu},
 time\rightarrow\frac{\mu \sigma_m^2}{\beta^2 \varepsilon
I_0^2}.
\end{eqnarray}
where $\beta=d/H<<1$ for a shallow  channel.

For a shallow channel, $\beta\rightarrow 0$ in the governing
equations which is the Hele-Shaw limit. This leads to
$\phi'=\phi'(x,y)$, $\sigma'=\sigma'(x,y)$, and
$\rho_b'=\rho_b'(x,y)$. It is assumed that $\partial\phi'/\partial
z = \partial\sigma'/\partial z = 0$ at the top and bottom walls
\cite{pata11a}.

In the Hele-Shaw limit the velocity component in the vertical
direction is zero, the pressure $p'=p'(x, y)$, and the horizontal
velocity in the $x-y$ plane is of the form
$\mathbf{u'}(x,y,z)=1.5(1-z^2)\mathbf{u}_m'(x,y)$
\cite{chen05a,linh08a,pata11a}, where the variables are
non-dimensional. As discussed earlier \cite{pata11a}
$\mathbf{u}_m'$ is the perturbation velocity at a given location
that is averaged with respect to the $z$ direction. The governing
equations become \cite{chen05a,linh08a,pata11a}
\begin{equation}
\left.
\begin{array}{l}
\nabla_H\cdot\mathbf{u}_m' = 0,\\
\nabla_H^2\phi' =-\rho_b', \\
\displaystyle\sigma_{0N}^2\nabla_H^2\phi'-\frac{\partial\sigma'}{\partial y} =0,\\
\displaystyle Pe\frac{\partial{\sigma}}{\partial{t}}=\nabla_H^2\sigma,\\
\displaystyle Re\beta^2\frac{\partial\mathbf{u}_m'}{\partial t}=
-\nabla
p'+\beta^2\nabla_H^2\mathbf{u}_m'-3\mathbf{u}_m'+\frac{\rho_b'}{\sigma_{ON}}\mathbf{j},\label{eqn:shallowndperturbation}
\end{array}
\right\}
\end{equation}
where $\nabla_H$ denotes the gradient in the $x-y$ plane. All
variables carry same meaning as before unless specified otherwise.
The non-dimensional parameters in this case are given by
\begin{equation}
\left.
\begin{array}{r}
\displaystyle Re=\frac{\beta^2 H^2 \rho\varepsilon I_0^2}{\mu^2\sigma_m^2},\\
\displaystyle Pe=\frac{\beta^2 H^2 \varepsilon I_0^2}{\mu D_\sigma \sigma_m^2},\\
\displaystyle\sigma_{0N}=\frac{2\sigma_0}{\sigma_0^{a}+\sigma_0^{b}}.
\end{array}
\right\}
\end{equation}

As discussed earlier \cite{pata11a}, the terms involving $\beta^2$
in the last of equation (\ref{eqn:shallowndperturbation}) should
be dropped in the Hele-Shaw limit. However, those terms may be
retained to approximately capture the viscous effects due to the
flow in the $x-y$ plane \cite{chen05a,linh08a}. Matching of the
``inner" solution in the thin viscous layers near the vertical
walls with the ``outer" Hele-Shaw solution is necessary to obtain
a formal solution. Such an analysis will not be considered here.
Instead, an approximate approach based on Equation
(\ref{eqn:shallowndperturbation}) will be considered
\cite{chen05a,linh08a}. This will also facilitate comparison with
the parallel electric field case considered earlier
\cite{pata11a}.

The novelty of our effort is the use of a sharp interface approach
in the linear stability analysis. Assuming perturbations of the
form given by Equation (\ref{eqn:solnperturb}) the governing
equations become
\begin{equation} \left.
\begin{array}{l}
\displaystyle(D^2-k^2-Pes)\sigma=0,\\
\displaystyle(D^2-k^2)\phi=\frac{D\sigma}{\sigma_{ON}^2},\\
\displaystyle(D^2-k^2-Res-\frac{3}{\beta^2})(D^2-k^2)v_m-\frac{k^2D\sigma}{\beta^2\sigma_{ON}^3},\\
\displaystyle u_m=-\frac{iDv_m }{k},\\
\displaystyle p=\frac{(\beta^2(D^2-k^2)-\beta^2s-3){Dv_m}}{k^2}.
\end{array}
\right\}
\end{equation}
Solutions of the governing equations are given by
\begin{equation}
\left.
\begin{array}{l}
\sigma^a=A^a\sinh{\lambda y}+B^a\cosh{\lambda y}\nonumber,\\
\displaystyle\phi^a=C^a\sinh{\lambda y}+D^a\cosh{\lambda
y}+\frac{D\sigma^a}{(\sigma_{0N}^a)^2(\lambda^2-k^2)}\nonumber ,\\
\displaystyle v^a=E^a\sinh{\lambda y}+F^a\cosh{\lambda
y}+G^a\sinh{\lambda y}+H^a\cosh{\lambda y}\nonumber\\
\hspace{1cm}-\displaystyle{k^2\frac{D\sigma^a}{\beta^2(\sigma_{0N}^a)^3(\lambda^2-q^2)(\lambda^2-k^2)}},\\
\end{array}
\right\}
\end{equation}

\begin{equation}
\left.
\begin{array}{l}
\sigma^b=A^b\sinh{\lambda y}+B^b\cosh{\lambda y}\nonumber,\\
\displaystyle\phi^b=C^b\sinh{\lambda y}+D^b\cosh{\lambda
y}+\frac{D\sigma^b}{(\sigma_{0N}^b)^2(\lambda^2-k^2)}\nonumber ,\\
\displaystyle v^b=E^b\sinh{\lambda y}+F^b\cosh{\lambda y}+G^b\sinh{\lambda y}+H^b\cosh{\lambda y}\nonumber\\ \hspace{1cm-}\displaystyle{k^2\frac{D\sigma^b}{\beta^2(\sigma_{0N}^b)^3(\lambda^2-q^2)(\lambda^2-k^2)}}. \\
\end{array}
\right\}
\end{equation}

Superscripts $a$ or $b$ denote liquids $a$ or $b$, respectively.
$A, B, C, D, E, F, G$ and $H$ are constants. $\lambda$ and $q$ are
positive and are given by
\begin{equation}
\left.
\begin{array}{l}
\lambda=\sqrt{k^2+Pes}\nonumber,\\
\displaystyle q=\sqrt{k^2+Res+\frac{3}{\beta^2}}.
\end{array}
\right\}
\end{equation}

Now we must use the boundary and interface conditions to solve for
the constants in the solution. The linearized jump conditions for
electrical conductivity at the interface are
\begin{equation}
\left.
\begin{array}{l}
\|\sigma\|=0 \nonumber,\\
\|D\sigma\|=Pev_{int}\sigma_{0N}.
\end{array}
\right\} \label{eqn:jcforconductivity}
\end{equation}
The linearized boundary conditions for electrical conductivity at
the side walls (i.e. at $y=\pm1$) are given by $D\sigma=0$. The
interface conditions together with the boundary conditions give
the following solution for the constants
\begin{equation}
\left.
\begin{array}{l}
\displaystyle A^{a}=-A^{b}=-\frac{Pe v_{\mbox{int}}\Delta\sigma_{0N}}{2\lambda}\nonumber,\\
\displaystyle B^{a}=B^{b}=-\frac{Pe
v_{\mbox{int}}\Delta\sigma_{0N}}{2\lambda\tanh\lambda},
\end{array}
\right\} \label{eqn:aandb}
\end{equation}
where $\Delta\sigma_{0N}=\sigma_{0N}^a-\sigma_{0N}^b$, and
$\displaystyle v_{\mbox{int}} = v_m^a = v_m^b =
\frac{v_m^a+v_m^b}{2}$ at $y=0$ is the $y$ component of velocity
at the interface.

The interface conditions for the electrical potential are
\begin{equation}
\left.
\begin{array}{l}
\|\phi\|=0\nonumber,\\
\displaystyle\|\sigma_{0N}D\phi\|+\sigma_{y=0}\frac{\Delta\sigma_{0N}}{\sigma_{0N}^+\sigma_{0N}^-}
\end{array}
\right\}\mbox{at y=0}.\label{eqn:jcforphi}
\end{equation}
Since there are electrodes at $y=\pm1$, there is no perturbation
of the electric potential at those boundaries. This implies
$\phi=0$ at $y=\pm1$. Using these interface and boundary
conditions we solve for $C^{a}$, $C^{b}$, $D^{a}$, and $D^{b}$ to
obtain
\begin{equation}
\left.
\begin{array}{l}
\displaystyle
{C^{a}=\frac{\Delta\sigma_{0N}}{4\sigma_{0N}^a\sigma_{0N}^b}\frac{k}{\lambda^2-k^2}(\frac{-1}{\tanh\lambda})\frac{Pev_{\mbox{int}}\Delta\sigma_{0N}}{\lambda}
+\frac{\mbox{Pe}v_{\mbox{int}}\Delta\sigma_{0N}\sigma_{0N}^b}{4(\lambda^2-k^2)\tanh
k}\frac{\sigma_{0N}^{a2}+\sigma_{0N}^{b2}}{\sigma_{0N}^{a2}\sigma_{0N}^{b2}}\nonumber},\\
\displaystyle
{C^{b}=\frac{\Delta\sigma_{0N}}{4\sigma_{0N}^a\sigma_{0N}^b}\frac{k}{\lambda^2-k^2}(\frac{1}{\tanh\lambda})\frac{Pev_{\mbox{int}}\Delta\sigma_{0N}}{\lambda}
+\frac{\mbox{Pe}v_{\mbox{int}}\Delta\sigma_{0N}\sigma_{0N}^a}{4(\lambda^2-k^2)\tanh
k}\frac{\sigma_{0N}^{a2}+\sigma_{0N}^{b2}}{\sigma_{0N}^{a2}\sigma_{0N}^{b2}}\nonumber},\\
D^{a}=-\tanh{k} C^{a}\nonumber,\\
D^{b}=\tanh{k} C^{b}.
\end{array}
\right\} \label{eqn:candd}
\end{equation}

The interface conditions for velocity are $\|Dv_m\|=0$ (which
follows from $\|u_m\|=0$) and $\|v_m\|=0$ at $y=0$. The velocity
boundary conditions are $v_m=0$ and $Dv_m=0$ (i.e. $u_m=0$) at
$y=\pm1$. The stress conditions at the interface are
$\displaystyle\|-p+2\beta^2Dv-\frac{D\phi}{\sigma_{0N}}\|=0$ and
$\displaystyle\|D^2v+v-\frac{\phi}{\sigma_{0N}}\|=0$. Using these
conditions together with Equations (\ref{eqn:aandb}) and
(\ref{eqn:candd}) we get eight equations for the constants $E$,
$F$, $G$ and $H$ in liquids $a$ and $b$. Only four of those
equations and the equation $\displaystyle
v_{\mbox{int}}=\frac{v_{m}^a+v_{m}^b}{2}$ give the maximum growth
rate:
\[
\left(
\begin{array}{ccccc}
 \sinh k & \cosh k & \sinh q & \cosh q & 0\\
 k\cosh k &  k\sinh k & q\cosh q & q\sinh q &
 \displaystyle{-\frac{k^2\lambda^2P_\Sigma}{2\lambda\sinh\lambda(\lambda^2-q^2)(\lambda^2-k^2)}}\\
 q & 0 & q & 0& \displaystyle{-\frac{k^2\lambda^2P_\Sigma}{2\lambda\tanh\lambda(\lambda^2-q^2)(\lambda^2-k^2)}}\\
 k^3 & 0 & q^3 & 0 &  f(k,\lambda,q,\Gamma,P_\Sigma)\\
 0 & 1 & 0 & 1&\displaystyle{(\frac{k^2P_\Sigma}{2(\lambda^2-k^2)(\lambda^2-q^2)-2})}
\end{array}
\right ) \left(
\begin{array}{l}
 E^s\\
 F^t\\
 G^s\\
 H^t\\
 v_{\mbox{int}}
\end{array}
\right )
\]
\begin{equation}
=\left (
\begin{array}{l}
 0\\
 0\\
 0\\
 0\\
 0
 \end{array}
 \right )\label{eqn:dispersion}
\end{equation}
In Equation (\ref{eqn:dispersion}), $E^s=E^a-E^b$ and
$F^t=F^a+F^b$. $G^s$ and $H^t$ are defined similarly.
$P_\Sigma=Pe\Sigma\Gamma/\beta^2$, where
$\displaystyle\Sigma=\frac{\Delta\sigma_{0N}^2}{\sigma_{0N}^{a2}\sigma_{0N}^{b2}}$
and
$\displaystyle\Gamma=\frac{\sigma_{0N}^a}{\sigma_{0N}^b}+\frac{\sigma_{0N}^b}{\sigma_{0N}^a}+1$.

The dispersion equation is once again obtained by setting the
determinant of the matrix, above, equal to zero.

\subsection{Results}

The dispersion equation gives $P_\Sigma$ as a function of
$\lambda,k,q, \mbox{and}\hspace{0.1cm} \Gamma$.
\begin{eqnarray}
P_\Sigma=-(32\Gamma\lambda(-k^2+\lambda^2)q(k^2-q^2)(\lambda^2-q^2)(k\mbox{cosh}q\mbox{sinh}k-q\mbox{cosh}k\mbox{sinh}q))\nonumber\\
/(k^2(\frac{1}{2}(-1+\Gamma)k\lambda q(\lambda^2-q^2)\mbox{coth}k(31+\mbox{cosh}k^2-16\mbox{cosh}k\mbox{cosh}q+\mbox{sinh}k^2)\nonumber\\
+8\mbox{coth}\lambda(q(-\Gamma\lambda^2q^2+k^2((-2+\Gamma)\lambda^2+2q^2))-q(-\Gamma\lambda^2q^2+k^2((-2+\Gamma)\lambda^2+2q^2))\nonumber\\
\mbox{cosh}k\mbox{cosh}q+k(k^2((-1+\Gamma)\lambda^2+q^2)+q^2((-1+\Gamma)\lambda^2+q^2))\mbox{sinh}k\mbox{sinh}q)+\nonumber\\
\lambda(-8q\mbox{cosh}q(-(-1+\Gamma)k(\lambda^2-q^2)\mbox{csch}k+\Gamma\lambda(k^2-q^2)\mbox{csch}\lambda+k(\Gamma k^2+\nonumber\\
\lambda^2-\Gamma\lambda^2-q^2)\mbox{sinh}k+\mbox{cosh}k(8\Gamma\lambda q(k^2-q^2)\mbox{csch}\lambda+(-1+\Gamma)kq(\lambda^2-q^2)\mbox{sinh}k-\nonumber\\
8(q^2((-1+\Gamma)\lambda^2+q^2)+k^2((-1+\Gamma)\lambda^2+(1-2\Gamma)q^2))\mbox{sinh}q))))\nonumber.\\
\end{eqnarray}
$P_\Sigma$ can also be expressed as a function of $\displaystyle
k,\mbox{Pes},(\mbox{Res}+\frac{3}{\beta^2}),
\mbox{and}\hspace{0.1cm}\Gamma$ i.e.
\begin{equation}
P_\Sigma=f(\lambda,k,q,\Gamma)=g(k,\mbox{Pes},(\mbox{Res}+\frac{3}{\beta^2}),\Gamma).
\end{equation}
For marginal stability at low Re we take $Re\rightarrow0$ and
$s\rightarrow0$. In this case $P_\Sigma$ depends on
$\Gamma,k,\mbox{and}\hspace{0.1cm}\beta$ i.e.
\begin{equation}
P_\Sigma=f(\Gamma,k,\beta).
\end{equation}

The trends of $P_\Sigma$ will be considered next. We will use
parameters corresponding to typical experimental values
\cite{chen05a}: $d=5.5\mu m$, $H=78\mu m$, $\rho=1000kg/m^3$,
$\mu=0.001kg/ms$, $\varepsilon=6.9\times 10^{-10}C/Vm$,
$D_{\sigma}=10^{-9}m^2/s$, and
$\displaystyle\frac{\sigma_0^a}{\sigma_0^b} = 10$. This implies
$\beta=0.07$, $\Gamma = 11.1$, and $\Sigma = 24.5$. The only free
variables are $I_0/\sigma_m$ and $k$. In this case, $P_\Sigma$ is
therefore the non-dimensional parameter that represents the
variation of $I_0/\sigma_m$ which is an average measure of the
external electric field.

\begin{figure}
\centering
\includegraphics[width=13cm]{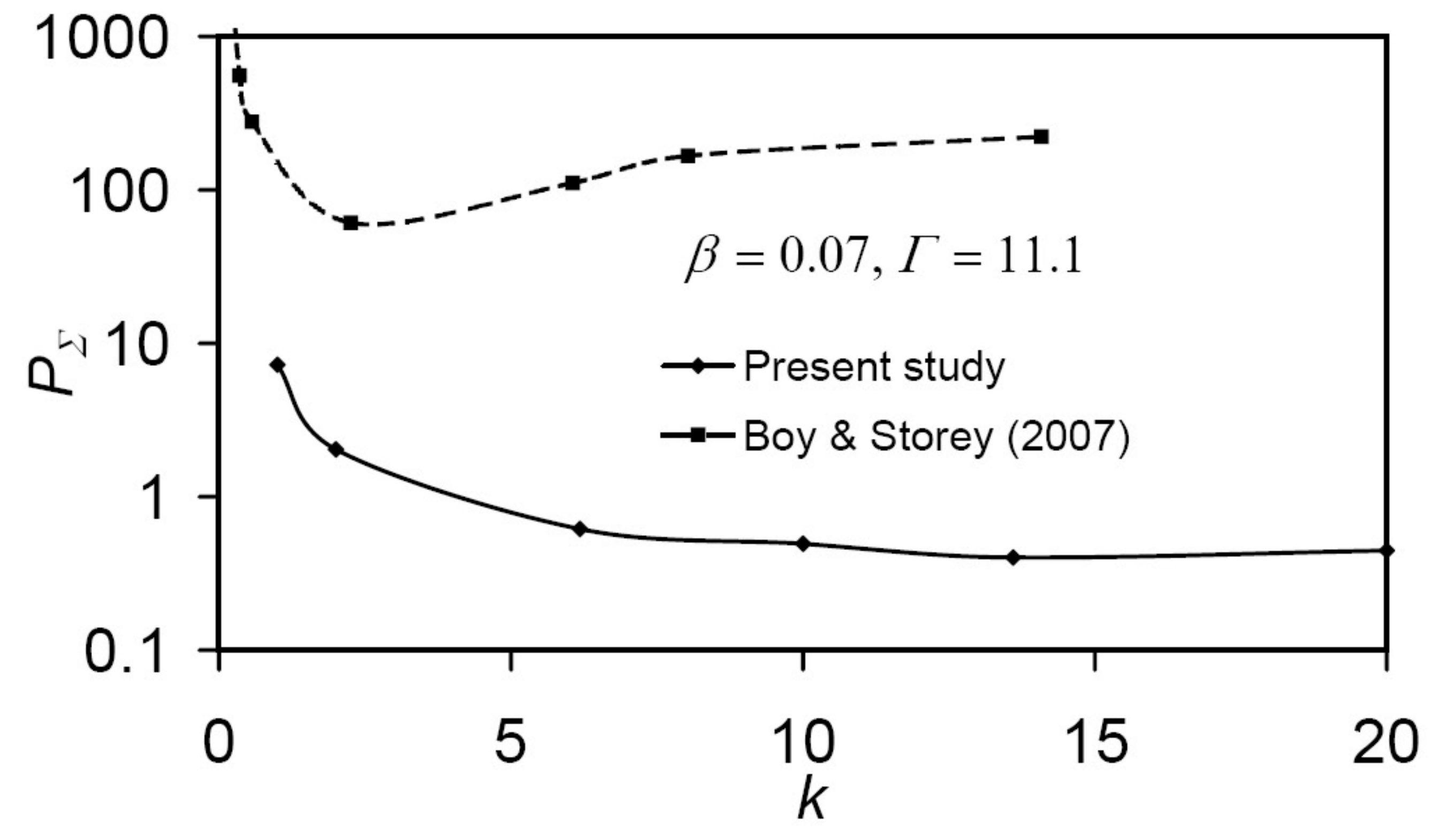}
\caption{The marginal stability curve of $P_\Sigma$ vs. $k$ for
$\beta$=0.07 and $\Gamma$=11.1.} \label{fig:psigmavsk}
\end{figure}

Figure (\ref{fig:psigmavsk}) shows the marginal stability curve of
$P_\Sigma$ vs.the wavenumber $k$ obtained from the dispersion
equation discussed above. It is seen that there is a critical
value of the $P_\Sigma$ (or correspondingly $I_0/\sigma_m$) below
which the system is stable. This is consistent with the threshold
type behavior seen in experiments \cite{oddy01a}. Figure
(\ref{fig:psigmavsk}) shows that the system becomes unstable at
$P_\Sigma^{crit} = 40211.4$. This implies a critical value of
$(I_0/\sigma_m)^{crit} = 0.06kV/cm$. Typical values in experiments
are $0.1-1kV/cm$ \cite{oddy01a,chen05a}. This suggests that an
instability should be observed under experimental conditions,
which is consistent with the data \cite{oddy01a}.

Figure (\ref{fig:psigmavsk}) also shows that at $P_\Sigma^{crit} =
40211.4$ the unstable wave corresponds to $k = 13.6$ which implies
a wavelength of the instability that is $0.23$ times the channel
width. Typical wavelengths of the instability are reported to be
of the order of the device width \cite{oddy01a}.

Boy and Storey \cite{boyd07a} presented an instability analysis
for the same configuration as that considered in this work with
the only difference being that they considered a diffuse
interface, in the base state, that was $0.2$ times the channel
width. In our work, we consider the limiting case of a sharp
interface. Figure (\ref{fig:psigmavsk}) shows a comparison of the
marginal stability curve from Boy and Storey \cite{boyd07a} and
the present analysis. It is seen that in the perpendicular
electric field case the diffusion at the interface can alter the
onset of instability significantly. The sharp interface limit sets
a lower bound on the critical condition. It is noted that the
electric Rayleigh number $Ra$ in the plot of Boy and Storey
\cite{boyd07a} is related to the parameter $P_\Sigma$ in this work
according to the following relation: $P_\Sigma=Ra\Sigma\Gamma$.
This relation is used to re-plot the data of Boy and Storey
\cite{boyd07a} in terms of $P_\Sigma$ in Figure
(\ref{fig:psigmavsk}).

The critical value for the onset of instability $P_\Sigma^{crit}$
identified in Figure (\ref{fig:psigmavsk}) depends on the the
conductivity ratio of the two liquids (i.e. on $\Gamma$) and also
on the channel height to width ratio (i.e. on $\beta$). This is
considered next.

\begin{figure}
\centering
\includegraphics[width=13cm]{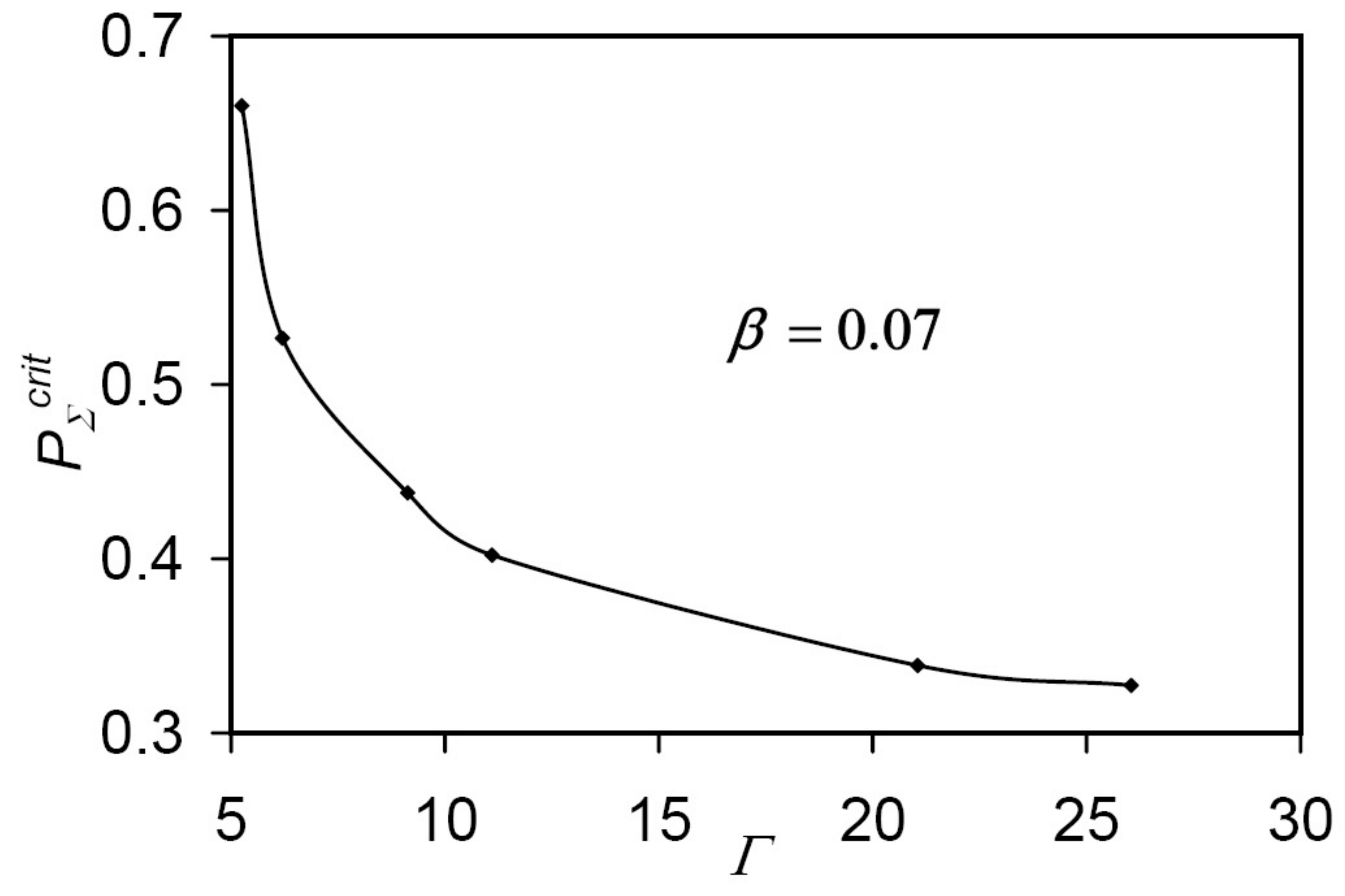}
\caption{Plot of $P_\Sigma^{cri}$ vs. $\Gamma$ indicating critical
conditions for the onset of instability at $\beta$=0.07.}
\label{fig:psigmcriavsgamma}
\end{figure}

\begin{figure}
\centering
\includegraphics[width=13cm]{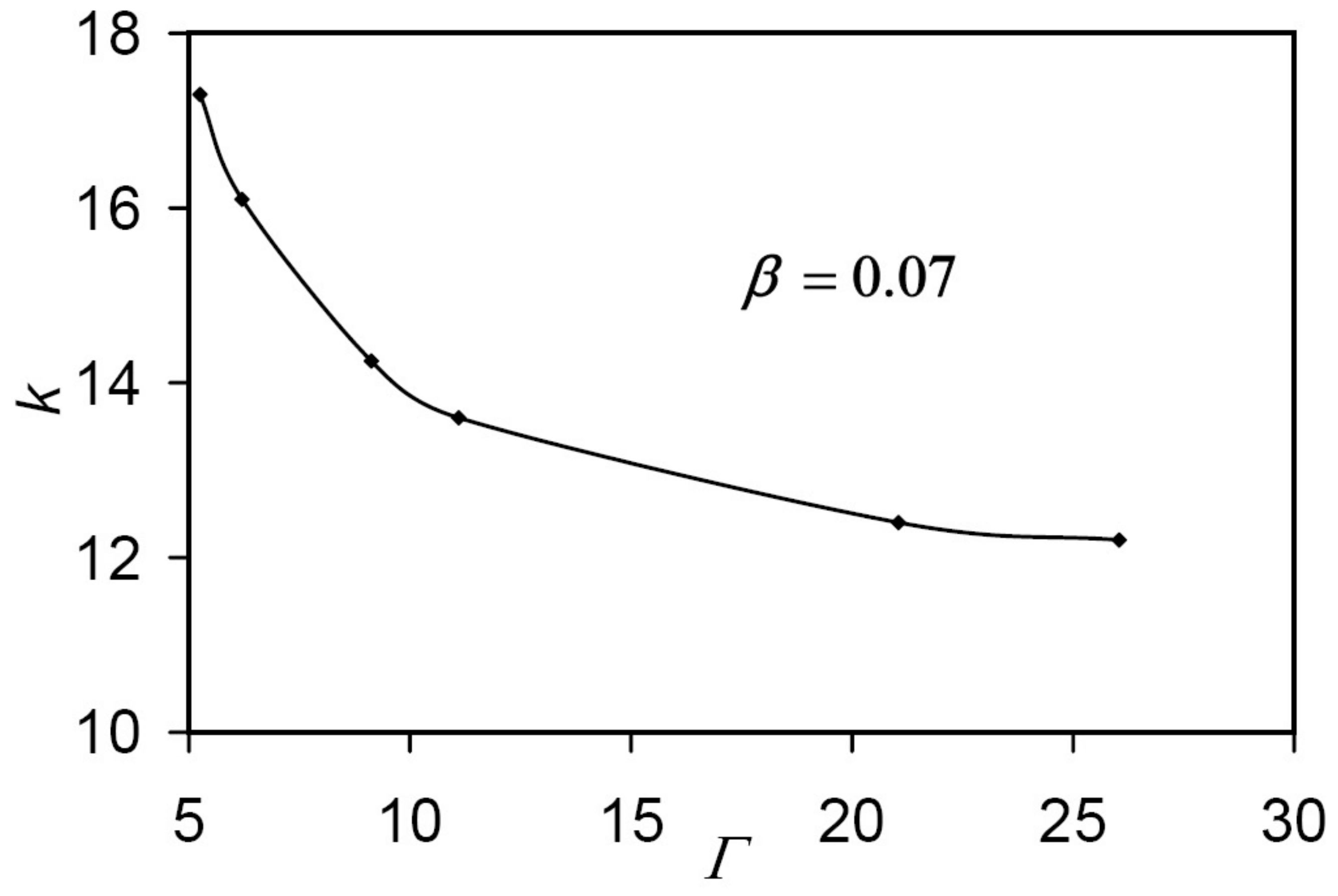}
\caption{Plot of $k$ vs. $\Gamma$ at critical conditions for the
onset of instability at $\beta$=0.07.} \label{fig:kvsgamma}
\end{figure}

\begin{figure}
\centering
\includegraphics[width=13cm]{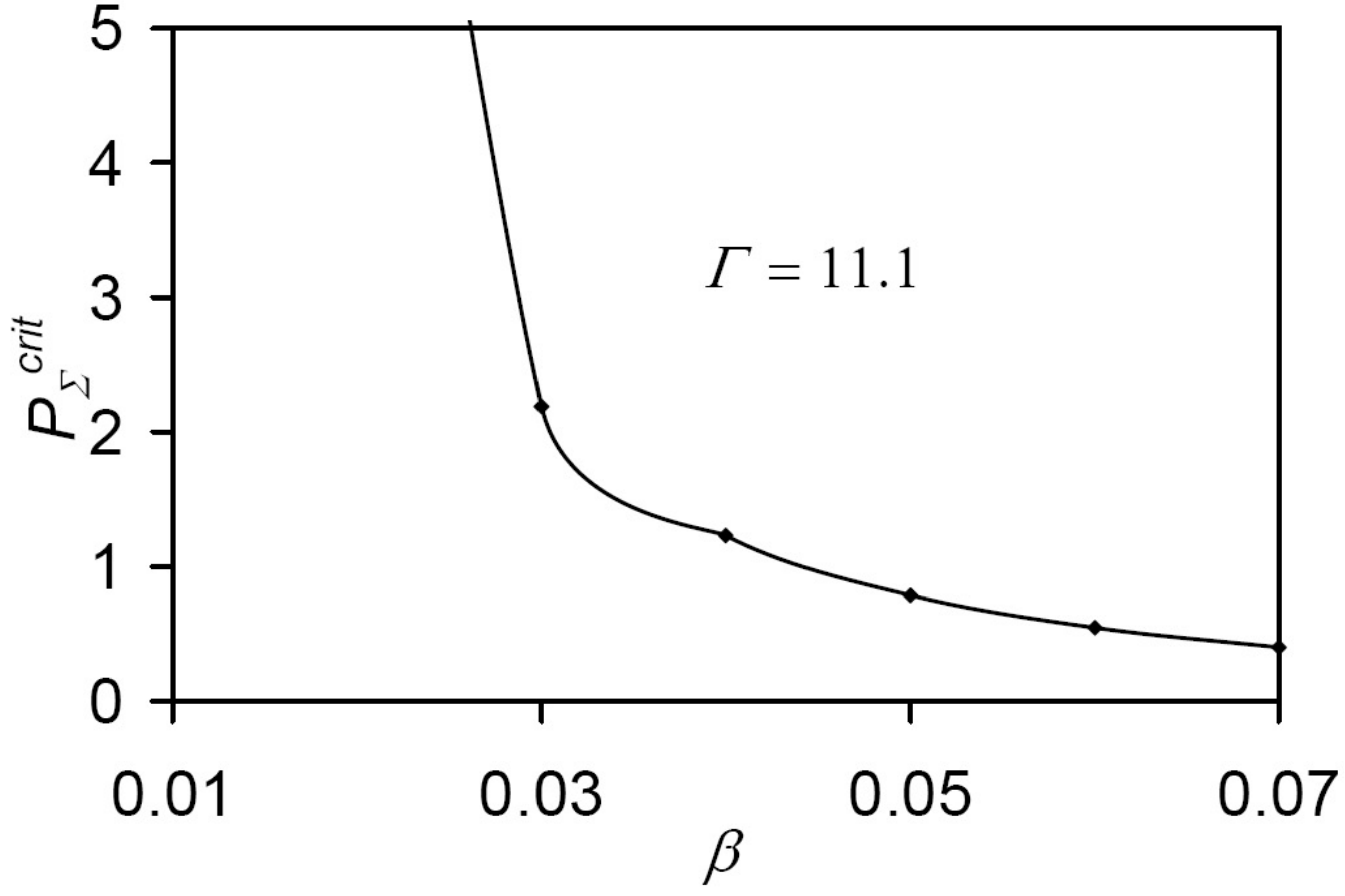}
\caption{Plot of $P_\Sigma^{cri}$ vs. $\beta$ indicating critical
conditions for the onset of instability at $\Gamma$=11.1.}
\label{fig:psigmavsbeta}
\end{figure}

\begin{figure}
\centering
\includegraphics[width=13cm]{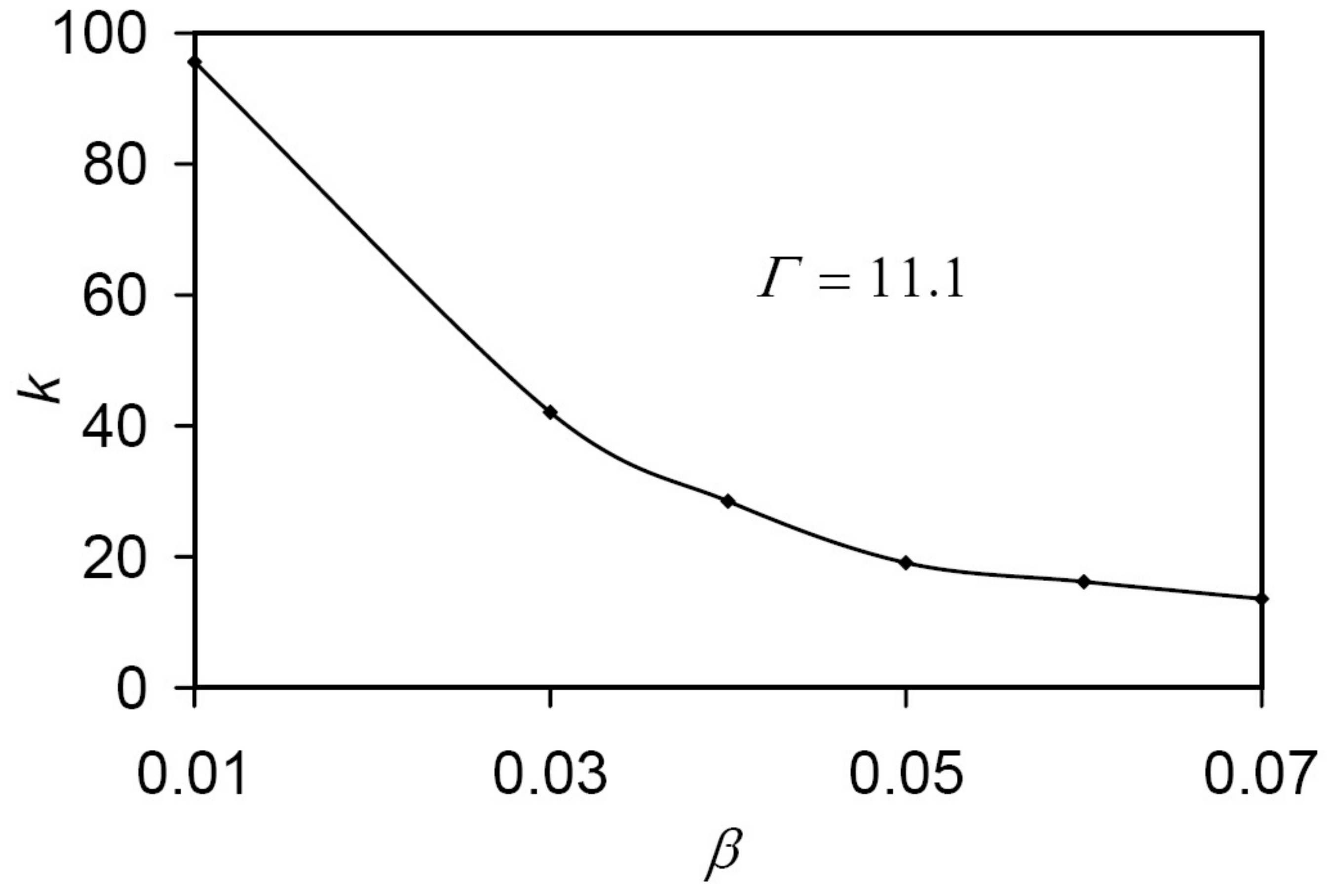}
\caption{Plot of $k$ vs. $\beta$ at critical conditions for the
onset of instability at $\Gamma$=11.1.} \label{fig:kvsbeta}
\end{figure}

Figure (\ref{fig:psigmcriavsgamma}) shows $P_\Sigma^{cri}$ vs.
$\Gamma$ indicating the critical condition for the onset of
instability at $\beta=0.07$. As expected, when $\Gamma$ increases
the critical value of $P_\Sigma$ decreases. It implies that the
system is more unstable with a larger conductivity ratio between
the two liquids. The wavenumbers at the critical condition for
instability, in this case, are plotted in Figure
(\ref{fig:kvsgamma}).

Figure (\ref{fig:psigmavsbeta}) shows $P_\Sigma^{cri}$ vs. $\beta$
indicating the critical condition for the onset of instability at
$\Gamma=11.1$. As $\beta$ decreases, i.e. as the channel becomes
more shallow, the critical value of $P_\Sigma$ increases. It
implies that the shallow nature of the channel has a stabilizing
effect on the instability. The wavenumbers at the critical
condition for instability, in this case, are plotted in Figure
(\ref{fig:kvsbeta}).

Finally, we compare the cases where the electric field is applied
parallel or perpendicular to the liquid-liquid interface. Using
the same parameters as those used in Figure (\ref{fig:psigmavsk}),
it was found that the critical electric field for the onset of
instability for the parallel electric field case is $0.08kV/cm$
\cite{pata11a}. Comparing this with the value of $0.06kV/cm$
computed above for the perpendicular electric field case, it is
implied that the perpendicular electric field case is more
unstable. Although we have compared these numbers at the
experimentally relevant condition, this trend continues at other
parameters.

\section{Discussion}

Some comments pertaining to the problem formulation and future
experiments are in order. These issues are discussed below.

In this work we have assumed that there is no electroosmotic flow.
Similar assumption has been made in the past by, e.g., Boy and
Storey \cite{boyd07a}. While electroosmotic flow can affect the
instability, it has been reported in prior work that the influence
is little when the ratio of electroosmotic to electroviscous
velocity is small (see Lin et al. \cite{linh04a} and Chen et al.
\cite{chen05a}). Instabilities caused by electroosmotic slip
velocity have been studied by others (see Boy and Storey
\cite{boyd07a} for references). As noted by Boy and Storey
\cite{boyd07a}, such instabilities do not rely upon bulk
conductivity gradient and therefore occur by a different mechanism
than the one considered in this work.

We have used a constant current condition in the base state with
respect to which a linear stability analysis is done. The
electrodes have a Dirichlet boundary condition which is similar to
that used in prior analytic work
\cite{linh04a,oddy05a,chen05a,stor05a,posn06a,boyd07a,linh08a}.
Our work does not consider the charging of the double layer at the
electrodes. Boy and Storey \cite{boyd07a} note that ``the double
layer capacitance acts as a high-pass RC filter on the electric
field in the bulk. Double layer charging simply adds another
mechanism to reduce the instability." Thus, our work helps
establish a baseline with respect to which the effects of
electrode charging can be studied in the future.

The analysis presented here is strictly valid only when the time
scale of the growth rate is shorter than the diffusion time scale.
Yet, this analysis is useful to understand the nature of the
instability and its domain of unstable behavior. This has been
discussed in our previous work (Patankar \cite{pata11a}), where
thresholds for validity have been shown. Similar discussion is
also reported by Boy and Storey \cite{boyd07a}.

In our analysis, the infinite domain case is considered as a
reference case which would be relevant when the channel is very
wide. In microfluidic scenarios, this may not be applicable.
Hence, we have considered the shallow channel configuration,
experiments for which can be set up according to the problem
definition in the paper. This is no different from the analyses
presented by Santiago and co-workers
\cite{linh04a,oddy05a,chen05a,stor05a,posn06a,boyd07a,linh08a}. El
Moctar et al. \cite{elmo03a} have reported similar experiments but
they have a square cross section channel instead of a shallow
channel. Thus, direct comparison with their data is not feasible.

\section{Conclusion}

In this paper the instability at the interface between two
miscible liquids with identical mechanical properties but
different electrical conductivities was analyzed in the presence
of a perpendicular electric field. Linear stability analysis was
done by considering a sharp interface between adjacent liquids in
the base state. This approach enabled an analytic solution for the
critical condition of the electrokinetic instability. It was seen
that the instability depends on a non-dimensional parameter
$P_\Sigma$ defined in the Equation (\ref{eqn:pcri}).

The mechanism of instability was analyzed. It was found that the
electrohydrodynamic coupling due to the interface condition for
the electrical conductivity and the electrical body force in the
fluid equations led to the instability.

The perpendicular electric field case is more unstable compared to
the parallel electric field case. The reason for this is the
greater asymmetry in the perpendicular field case that results in
larger destabilizing electrohydrodynamic force.

The effect of a microchannel geometry was studied and the relevant
parameters were found to be $P_\Sigma, \beta$, and $\Gamma$ as
defined in the paper.

The analysis captured the threshold type behavior for the onset of
instability. It showed that larger conductivity ratio has a
destabilizing effect, while the shallow nature of the channel has
a stabilizing effect on the instability. Our approach provides a
theoretical estimate and scaling for the desired parameters.

\bibliographystyle{unsrt}

\bibliography{EKIRef}

\begin{thebibliography}{10}

\bibitem{pata11a}
NA~Patankar.
\newblock {Electrokinetic instability: The sharp interface limit}.
\newblock {\em {PHYSICS OF FLUIDS}}, {23}({1}):{014101}, {JAN} {2011}.

\bibitem{oddy01a}
MH~Oddy, JG~Santiago, and JC~Mikkelsen.
\newblock {Electrokinetic instability micromixing}.
\newblock {\em {ANALYTICAL CHEMISTRY}}, {73}({24}):{5822--5832}, {DEC 15}
  {2001}.

\bibitem{linh04a}
H~Lin, BD~Storey, MH~Oddy, CH~Chen, and JG~Santiago.
\newblock {Instability of electrokinetic microchannel flows with conductivity
  gradients}.
\newblock {\em {PHYSICS OF FLUIDS}}, {16}({6}):{1922--1935}, {JUN} {2004}.

\bibitem{oddy05a}
MH~Oddy and JG~Santiago.
\newblock {Multiple-species model for electrokinetic instability}.
\newblock {\em {PHYSICS OF FLUIDS}}, {17}({6}):{064108}, {JUN} {2005}.

\bibitem{chen05a}
CH~Chen, H~Lin, SK~Lele, and JG~Santiago.
\newblock {Convective and absolute electrokinetic instability with conductivity
  gradients}.
\newblock {\em {JOURNAL OF FLUID MECHANICS}}, {524}:{263--303}, {FEB 10}
  {2005}.

\bibitem{stor05a}
BD~Storey, BS~Tilley, H~Lin, and JG~Santiago.
\newblock {Electrokinetic instabilities in thin microchannels}.
\newblock {\em {PHYSICS OF FLUIDS}}, {17}({1}):{018103}, {JAN} {2005}.

\bibitem{posn06a}
JD~Posner and JG~Santiago.
\newblock {Convective instability of electrokinetic flows in a cross-shaped
  microchannel}.
\newblock {\em {JOURNAL OF FLUID MECHANICS}}, {555}:{1--42}, {MAY 25} {2006}.

\bibitem{linh08a}
H~Lin, BD~Storey, and JG~Santiago.
\newblock {A depth-averaged electrokinetic flow model for shallow
  microchannels}.
\newblock {\em {JOURNAL OF FLUID MECHANICS}}, {608}:{43--70}, {AUG 10} {2008}.

\bibitem{uguz08a}
A.~Kerem Uguz, O.~Ozen, and N.~Aubry.
\newblock {Electric field effect on a two-fluid interface instability in
  channel flow for fast electric times}.
\newblock {\em {PHYSICS OF FLUIDS}}, {20}({3}):{031702}, {MAR} {2008}.

\bibitem{uguz08b}
A.~Kerem Uguz and N.~Aubry.
\newblock {Quantifying the linear stability of a flowing electrified two-fluid
  layer in a channel for fast electric times for normal and parallel electric
  fields}.
\newblock {\em {PHYSICS OF FLUIDS}}, {20}({9}):{092103}, {SEP} {2008}.

\bibitem{boyd07a}
DA~Boy and Storey BD.
\newblock {Electrohydrodynamic instabilities in microchannels with time
  periodic forcing}.
\newblock {\em {PHYSICS REVIEW E}}, {76}:{026304}, {2007}.

\bibitem{Melcher}
J.~R. Melcher.
\newblock {\em Continuum Electromechanics}.
\newblock MIT Press, 1981.

\bibitem{elmo03a}
El~Moctar AO, Aubry N, and Batton J.
\newblock {Electro-hydrodynamic micro-fluidic mixer}.
\newblock {\em {LAB CHIP}}, {3}:{273--280}, {2003}.

\end{thebibliography}

\end{document}